\documentclass[aip,pof,reprint,groupedaddress,superscriptaddress]{revtex4-1}

\usepackage{graphicx}
\usepackage{subfigure}
\usepackage{hyperref}
\usepackage{amsmath}
\usepackage{bm}
\usepackage{color}

\graphicspath{{./Figures/}}	% COMMENTED for final submitted version!!

\newcommand{\bX}{\bm X}
\newcommand{\bu}{\bm u}

\newcommand{\St}{St}
\newcommand{\tp}{\tau_\mathrm{p}}

\newcommand{\new}[1]{{\textcolor{black}{#1}}}

\newcommand{\obs}{Laboratoire J.-L. Lagrange, Universit\'e de
  Nice-Sophia Antipolis, CNRS, Observatoire de la C\^ote d'Azur, 06300
  Nice, France} \newcommand{\mpids}{Max Planck Institute for Dynamics
  and Self-Organization (MPIDS), 37077 G\"ottingen, Germany}

\begin{document}

%%%
\title{Extreme fluctuations of the relative velocities between
  droplets in turbulent airflow}

\author{Ewe-Wei Saw}
\affiliation{\mpids}
\affiliation{\obs}

\author{Gregory P.\ Bewley}
\affiliation{\mpids} 
 
\author{Eberhard Bodenschatz}
\affiliation{\mpids}
\altaffiliation[]{Institute for Nonlinear Dynamics, University of
  G\"ottingen, G\"ottingen, Germany}
\altaffiliation[]{Lab. of Atomic \& Solid-State Phys. and Sibley
  School of Mech. \& Aerospace Eng., Cornell University, USA}

\author{Samriddhi Sankar Ray}
\affiliation{International Centre for Theoretical Sciences, Tata
  Institute of Fundamental Research, Bangalore 560012, India}

\author{J\'er\'emie Bec}
\affiliation{\obs}

\begin{abstract}

  We compare experiments and direct numerical simulations to evaluate
  the accuracy of the Stokes-drag model, which is used widely in
  studies of inertial particles in turbulence.  We focus on statistics
  at the dissipation scale and on extreme values of relative particle
  velocities for moderately inertial particles ($St < 1$).  The
  probability distributions of relative velocities in the simulations
  were qualitatively similar to those in the experiments.  The
  agreement improved with increasing Stokes number and decreasing
  relative velocity.  Simulations underestimated the probability of
  extreme events, which suggests that the Stokes drag model misses
  important dynamics.  Nevertheless, the scaling behavior of the
  extreme events in both the experiments and the simulations can be
  captured by the same multi-fractal model.
%(100 word limit)

\end{abstract}

\date{\today}

\maketitle

\noindent In warm clouds (with no ice), air-turbulence enhances the
collision rate of the droplets.  It thus influences the evolution of
droplet sizes and the timescale for rain
formation.\cite{falkovich2002acceleration,shaw2003particle} Two
mechanisms are at play: preferential concentration, due to a
combination of dissipative dynamics and non-trivial correlations
between the fluid flow and particle
positions,\cite{balkovsky2001intermittent,bec2007heavy,Saw2008} and
very large approach velocities, explained in terms of the {\it sling
  effect}\cite{falkovich2002acceleration,Bewley2013} and the formation
of {\it caustics}.\cite{wilkinson2006caustic,falkovich2007sling} Many
questions remain open regarding the impact of such phenomena on the
coalescence rate of droplets.  Whilst it is generally accepted that
turbulence increases droplet collision rates, too violent events can
cause fragmentation.\cite{orme1997experiments} To produce reliable
models for coalescence efficiencies, a key issue is to understand how
often this occurs.  Such considerations are decisive for unravelling
the impact of turbulence on the size distribution of droplets in
clouds.

Contemporary theories and simulations of heavy particle dynamics in
turbulent flows predominantly assume point particles coupled to the
flow through linear Stokes drag.  This simplification is
justified when the particles are (a) smaller than the smallest scales
of the flow, (b) made of material much denser than the fluid
(i.e. heavy), and (c) far apart.  Clearly the last premise fails
when particles come close enough to collide and subject to mutual
hydrodynamics interactions.  In addition, several corrections to
Stokes drag are missing from this framework and it is unclear when
they are needed to capture the full dynamics.  These include the
Basset history force, nonlinear drag and the added mass effect.
%In addition to hydrodynamic interactions between nearby particles, the corrections
%include the Basset history force, nonlinear drag and the added mass
%effect.  
Recent studies suggest that the history force tends to
suppress preferential concentration and caustic
formation.\cite{Hill2005,Daitche2011} To find out the extent to which
a model with Stokes drag alone is quantitatively descriptive,
we compare experiments of droplets in turbulent air flow to
results from direct numerical simulations (DNS) that match the
conditions of the experiment, but with point particles coupled to the
flow through Stokes drag.  We then investigate the scaling of the
particles' relative velocities with respect to their spatial
separation.  This scaling is relevant for predicting collisional
velocities at small scales from the large-scale statistics that are
more easily measured.  Finally, we compare our data with recent
theoretical results and investigate the nature of the transition from
tracer-like statistics at low relative velocities to the
particle-inertia dominated statistics at large relative velocities.

The experiment is described in detail in Ref.~\onlinecite{Bewley2013},
and only an overview is given here.  Nearly homogeneous and isotropic
turbulent flows are generated in a $1\;m$-diameter acrylic sphere by
32 randomly pulsating jets.  Each jet is made up of an audio-speaker
capped by a conical nozzle.\cite{Chang2012} The homogeneous and
isotropic region was about $10\;cm$ in diameter and at the center of
the apparatus.  We ran the experiment under three different
conditions, with the Taylor micro-scale Reynolds numbers,
$R_{\lambda}$, being 160, 170 and 190 and kinetic energy dissipation
rates ($\varepsilon$) $0.45 \pm 0.05$, $1.2 \pm 0.1$ and $3.2 \pm
0.2\;m^2/s^3$, respectively (the corresponding Kolmogorov dissipative
micro-scales, $\eta$, were 300, 230 and $180\;\mu m$).  Droplets are
produced with a spinning disc device\cite{walton49} that eject
bi-disperse drops with diameters $6.8\mu m$ and 19$\mu m$ and standard
deviations of $2\;\mu m$ and $4\;\mu m$, respectively.  The Stokes
numbers for the droplets are defined with respect to the Kolmogorov
time-scale as $\St=\tau_\mathrm{p}/\tau_{\eta}$ where $\tau_{\eta} =
\sqrt{\nu/\varepsilon}$ is the Kolmogorov timescale and $\tp =
(2/9)(\rho_\mathrm{p}/\rho_\mathrm{f}) a^2/\nu$ is the particle
viscous response time ($\rho_\mathrm{p}$ and $\rho_\mathrm{f}$ are the
particle and the fluid densities, respectively, $a$ the particle
radius and $\nu$ the fluid kinematic viscosity).  In order of
increasing $R_{\lambda}$ for the flows studied, the large (small)
droplets have Stokes number of values 0.19 (0.02), 0.31 (0.04) and
0.51 (0.06). Droplet motions are measured by imaging their shadows
projected by white light sources into two cameras fitted with macro
lenses, at a frame-rate of 15kHz ($>30/\tau_{\eta}$) and a spatial
resolution of $3\;\mu m/pixel$ ($<\eta/50$, \new{such unprecedented
  resolution allows us to measure the size and to distinguish the two
  groups of droplets}).  The 3D positions of droplets are determined
by stereoscopic Lagrangian Particle
Tracking.\cite{ouellette2006experimental}

The DNS are performed by using a
pseudo-spectral\cite{gottlieb1977numerical} parallel solver for the
fluid velocity $\bu$ obtained from the incompressible Navier--Stokes
equation.  Turbulence was sustained in a statistically stationary
regime by holding constant the energy content of the lowest Fourier
modes.\cite{chen1992high} We use $512^3$ grid points with $\nu =
1.5\times10^{-4}$ (corresponding to $R_{\lambda}=180$) to
approximately match the Reynolds numbers of the experiments.  The
droplets are approximated by individual point particles whose
trajectories $\bX(t)$ solve the Stokes equation
\begin{equation}
  \ddot{\bX} = -(1/\tp)\left[\dot{\bX} - \bu(\bX,t)\right] + \bm g, \label{eq:stokes}
\end{equation}
where the dots are time derivatives and $\bm g$ the acceleration of
gravity.  The fluid velocity at each particle position is obtained by
cubic interpolation from the grid points.  The point-particle
approach~\eqref{eq:stokes} is expected to be valid when the particles
size is much smaller than $\eta$ and their Reynolds number much less
than unity.  Furthermore, the particles in this model do not modify or
perturb the flow, which may be valid when their volume fraction is
small.

\begin{figure}
  \begin{center}
    %\subfigure[]{\includegraphics[width=.42\textwidth]{pdf_dv_St_0p05.pdf}}
    %\subfigure[]{\includegraphics[width=.42\textwidth]{pdf_dv_St_0p19.pdf}}\\ 
    %\subfigure[]{\includegraphics[width=.42\textwidth]{pdf_dv_St_0p31.pdf}}
    %\subfigure[]{\includegraphics[width=.42\textwidth]{pdf_dv_St_0p51.pdf}}
    \vspace{-50pt} {\includegraphics[width=.42\textwidth]{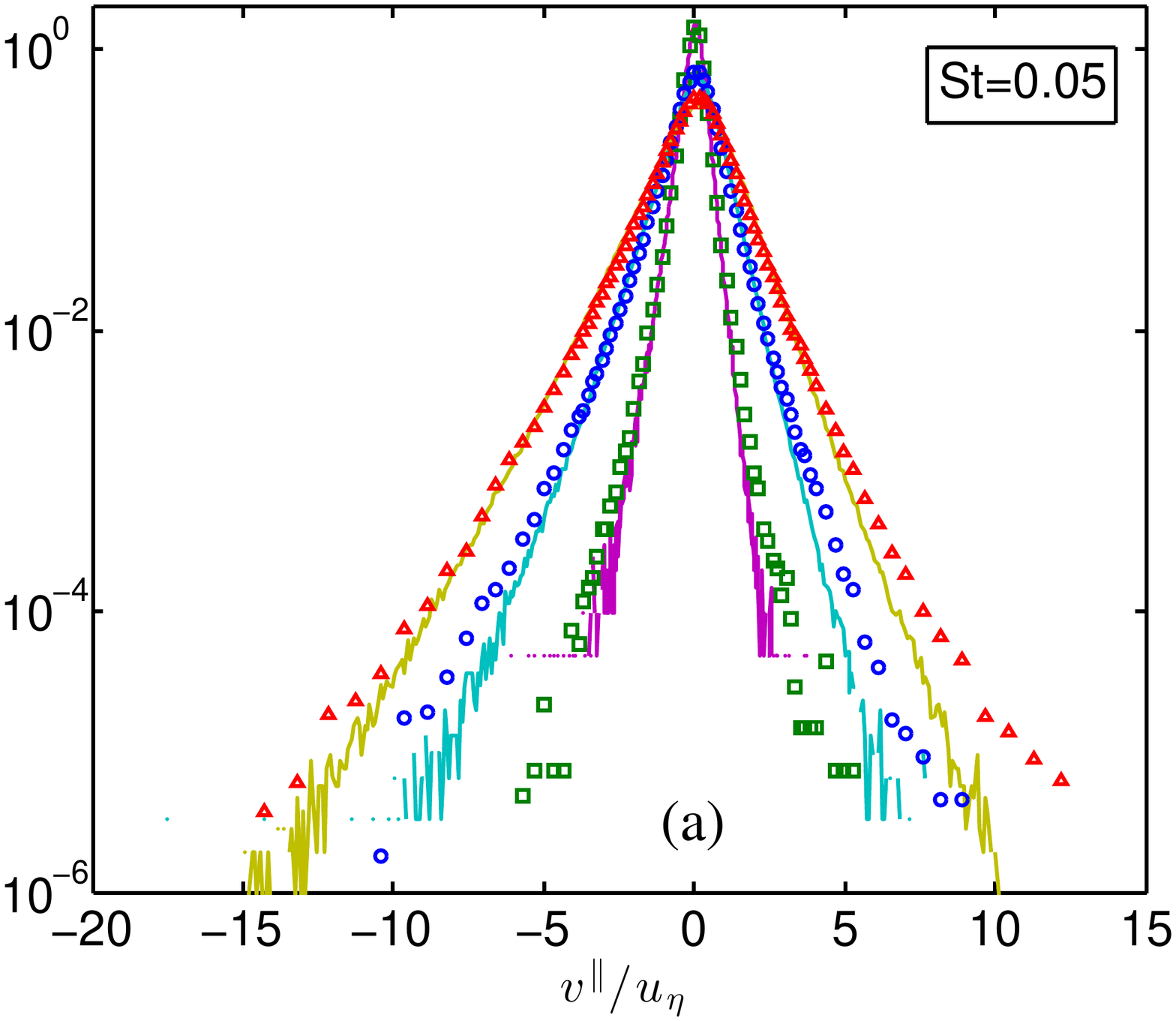}}  \vspace{-70pt}
    {\includegraphics[width=.42\textwidth]{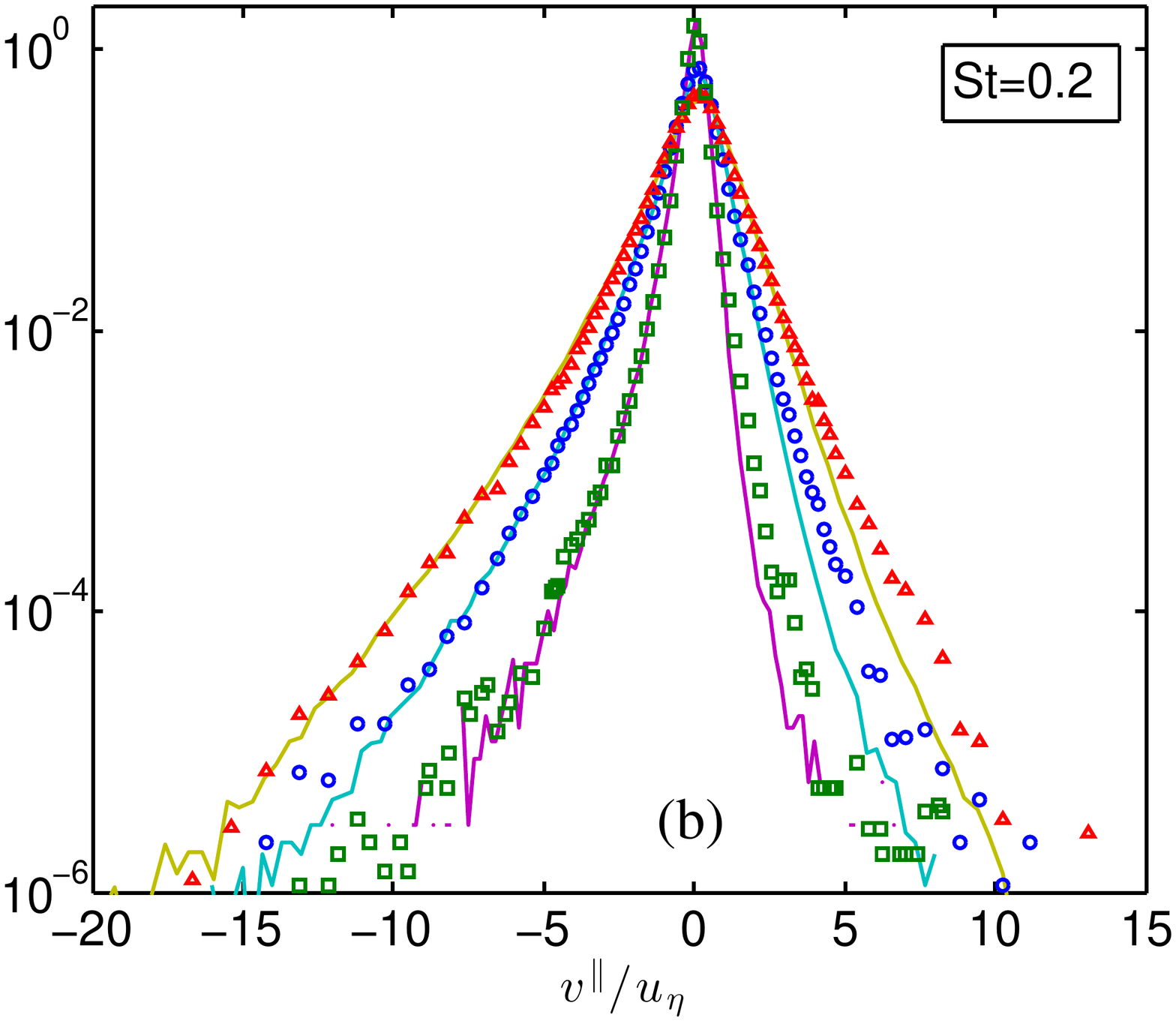}}  \\ 
    {\includegraphics[width=.42\textwidth]{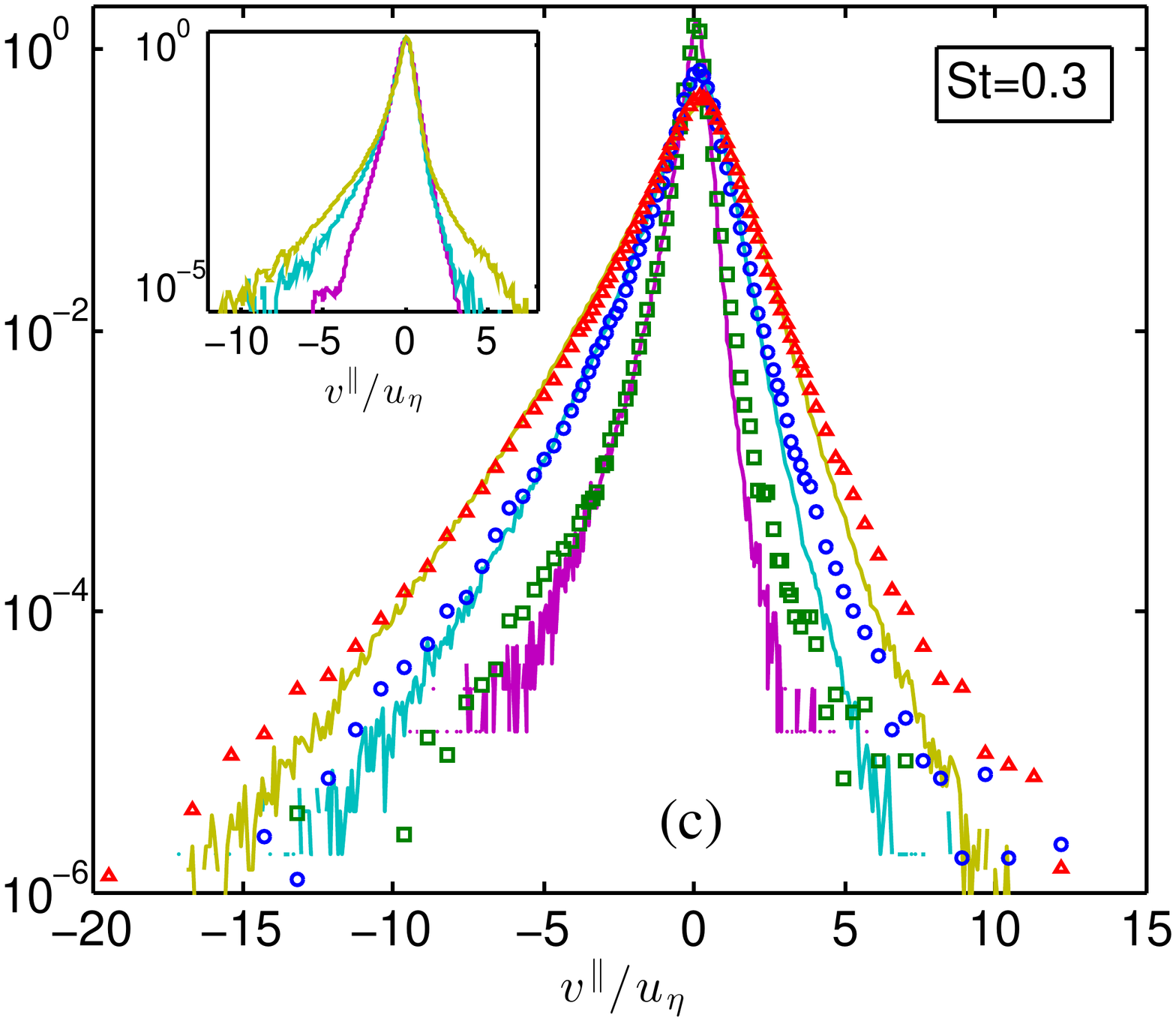}}
    {\includegraphics[width=.42\textwidth]{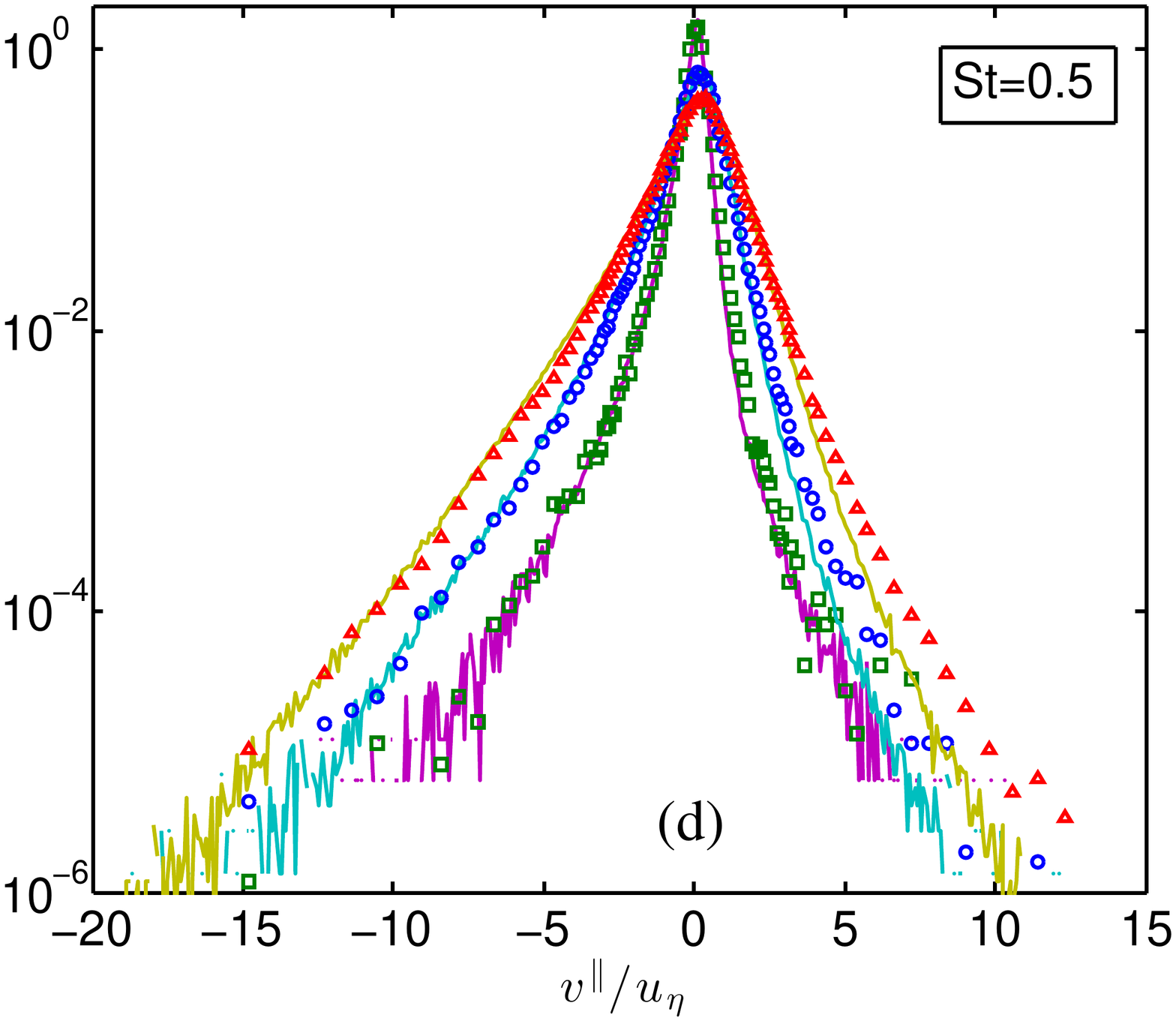}}
  \end{center}
\vspace{-50pt}
\caption{\label{Pdf_dv} (Color online) PDF of the longitudinal
  velocity differences conditioned on different separations $r$ for
  (a) $\St=0.05$, (b) $\St=0.2$ , (c) $\St= 0.3$ and (d) $\St=0.5$.
  Symbols are from experiments and solid lines from DNS.
  Squares/purple correspond to $r=1-1.6\eta$, circles/cyan to
  $r=3-3.6\eta$, and triangles/gold to $r=5-5.6\eta$.  \new{Note that unlike the DNS,
    the experimental droplets are not perfectly mono-disperse (details in
    text)}.  Inset of (c): PDF for $r=1-1.6\eta$ and from bottom to
  top: $\St=0.05$, $0.3$, $0.5$.}
\end{figure}

Of fundamental importance to the problem of turbulence-induced
collisions between particles are the statistics of their longitudinal
velocity difference $v^\parallel$ when the particles are close to each
other.  In Fig.~\ref{Pdf_dv} we show the probability density function
(PDF) of $v^\parallel$ between two particles, conditioned on different
values, $r$, of their separation.  The plots are organized into four
Stokes number groups: $\St = 0.05, 0.2, 0.3$ and $0.5$.  For some of
them, the experimental and simulated Stokes numbers slightly differ
(for $St = 0.05$, the experimental value is 0.04 and for $St = 0.2$,
the experimental and DNS values are 0.19 and 0.24 respectively).
\new{In the experiment the value of $R_{\lambda}$ changes a little
  between the various $St$ groups; thus minor Reynolds-number effect
  may be present.} There is general agreement in the trends and shapes
of the distributions.  All can be approximated by
stretched-exponentials whose concavity grows more pronounced with
increasing $\St$ and decreasing $r$.  This is qualitatively consistent
with what is known about the velocity distributions of fluid
particles, which grow more stretched with decreasing
scale.\cite{kailasnath1992probability}
% It is relevant to note that Ref.~\onlinecite{gustavsson2008variable}
% predicts \textit{compressed} exponential distributions for very
% large Stokes numbers.  The distributions we measure are stretched
% rather than compressed, and the implication is that the large $\St$
% limit taken in the theory does not accurately describe the
% intermediate $\St$ dynamics studied here.

Both experiments and DNS show an increase in the amplitude of the left
tail with increasing $\St$, manifest in an increased skewness (\new{more
  clearly in the inset of Fig.~\ref{Pdf_dv}c}).  This implies that
particles with larger inertia approach one another more violently on
average than lighter ones.  This is consistent
with the sling effect, where inertial particles fly towards each other
with relative velocities much higher than that of the background
fluid, as \new{explained in
  Ref.~\onlinecite{gustavsson2011distribution,Gustavsson2013}} and also
observed in Ref.~\onlinecite{Bewley2013}.  The faster approach should
enhance their collision rate.  \new{Although similar skewness is well
  documented for fluid tracers, here we show that the skewness is
  further enhanced by particle inertia over the range of scales observed.} The mechanism of this
  enhancement essentially involves occurrence of slings and subsequent
  damping by viscous drag. As seen in the inset, the advection
dominated cores of the PDF do not change while the tails
grow wider with increasing $St$, which makes the PDF more
concave than that of fluid tracers.  This observation is consistent
with the existence of a velocity scale $\sim r/\tau_{\rm p}$ that separates the
fluid-advection-dominated core of the PDFs from the inertia-dominated
tails.\cite{Bewley2013}

Quantitatively, we found the differences between experiments and
simulations to be less than about $15\%$ in the core of the
distributions.  Similarly, we found excellent agreement in the tails
of the distributions, but only for the largest Stokes number
($\St=0.5$), the smallest scale ($r<2\,\eta$), and for the left side
of the distributions corresponding to approaching particle pairs.  In
other cases, the experimental tails of the PDFs increasingly deviate
from the simulated ones as one moves to higher relative velocities.
The discrepancy is larger in the right tails, corresponding to
separating pairs, where in worst case the experimental data is
about 5 times the DNS data.  In the left tails, the
discrepancy is less severe, but worsens with decreasing $\St$, with
the largest discrepancy at a factor of two.

In the case of $\St=0.5$ (Fig.~\ref{Pdf_dv}d), the discrepancy in the
right tails seems at first glance to contradict the good agreement
observed for the left tails.  Here, effects beyond linear Stokes drag
maybe at play (e.g., the Basset history force, the added mass and
nonlinear drag forces).  For example, there is some indication in
recent numerical simulations that the history force plays an important
role under some conditions.\cite{Daitche2011} In any case, we could
not find a clear explanation for the discrepancies, despite
considering several possibilities including measurement uncertainty.
To capture its influence, we characterized the measurement noise and
added it to the DNS data.  This however resulted only in a negligible
widening of the tails of the distributions (the r.m.s.\ of the noise
was in the data about $10\%$ of $v^{\parallel}$ ).
% >> I didn't understand the comment in parentheses - is this what you meant?  
We also evaluated the accuracy of the method used to estimate
$\varepsilon$, in the experiment by applying the same method to the
DNS data i.e., by using $\left\langle [v^{\parallel}] ^2 \right\rangle
= \varepsilon\, r^2/(15\nu)$ for $r<5\eta$ and on particles of
$\St=0.05$).  This resulted in very good agreement (within 5\%) with
the direct measure of $\varepsilon$ in DNS, and so gave a strong support
to the $\varepsilon$ reported in the experiment.  \new{ We checked
  that Reynolds number effect could not account for the discrepancies by comparing
  DNS data at increasing Reynolds numbers (up to $R_\lambda = 287$).}
This addressed partially the question of small scale universality of
the turbulence statistics in the flows studied.  We also explored the
possibility of inaccuracy of $\nu$ in the experiment by reprocessing
the experimental data with a modified $\nu$ ($\pm30\%$) and found no
clear improvement.  The droplets' Reynolds numbers ($u_{\eta}a/\nu$)
were of the order of 0.1 on average, so that the effect of non-linear
drag on the droplets were typically negligible.
% >> what about during extreme events?  what was the reynolds number then?  
We note that given the conditions of our experiment, 
and specifically since $a/\eta$ was of the order of 0.1, 
the history force term stands next to the Stokes drag in the hierarchy of importance amongst 
the various forces on the droplets.\cite{Maxey83}  
In summary, 
the influences of nonlinear forces, hydrodynamic interactions, 
and non-universal turbulence statistics merit further study.  

\begin{figure}
  %\subfigure[]
  \vspace{-50pt} {\includegraphics[width=.42\textwidth]{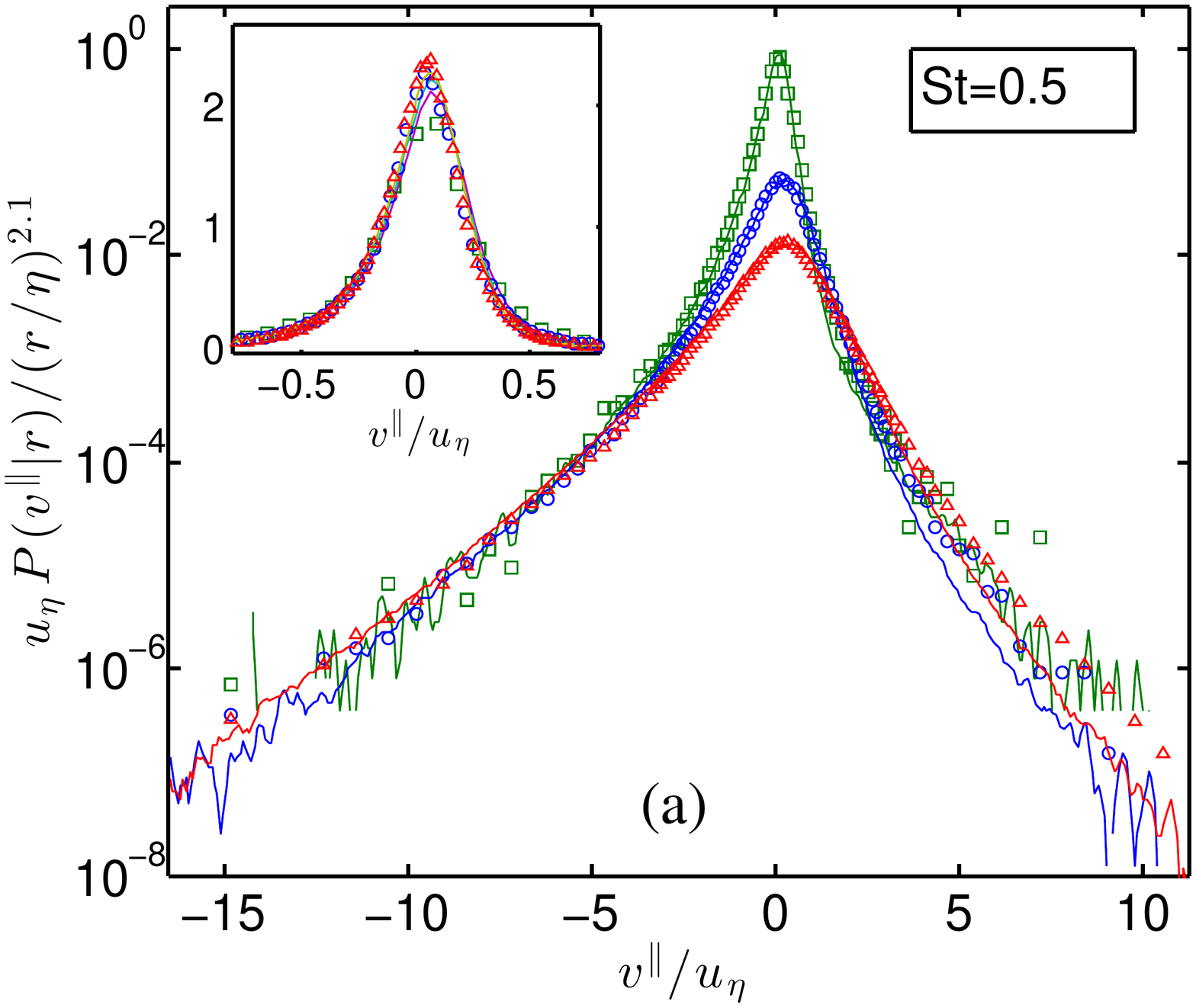}} 
    %\subfigure[]
  \
  {\includegraphics[width=.42\textwidth]{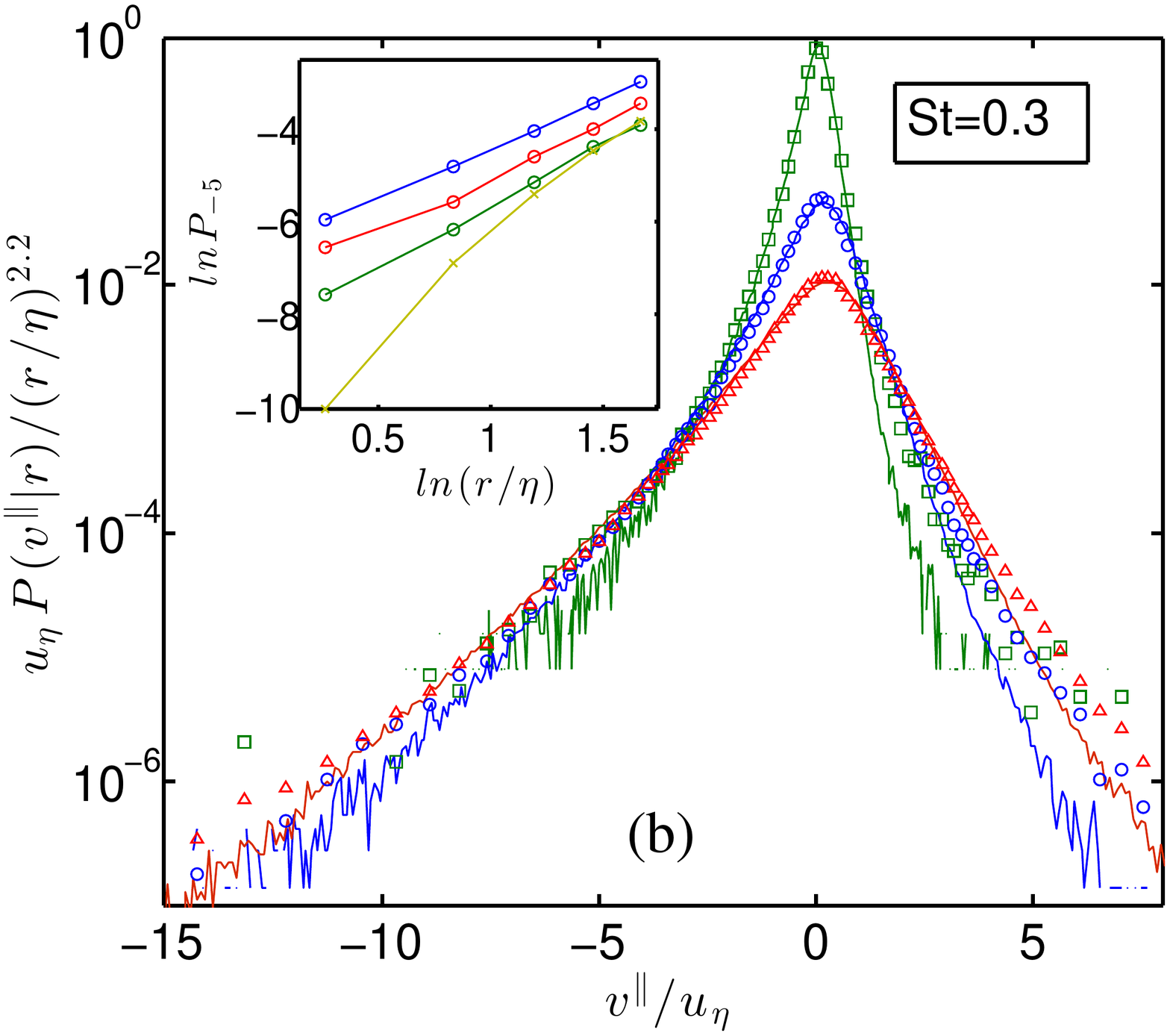}}  
  \vspace{-40pt}
  \caption{\label{fig:jointpdfs} (Color online) (a) Rescaled PDFs of
    the longitudinal velocity difference from the experiment (symbols)
    and the DNS (solid lines) for $\St=0.5$, with $\beta$ = 2.1, and
    different separations: $r=1-1.6\eta$ (green), $r=3-3.6\eta$
    (blue), and $r=5-5.6\eta$ (red).  Inset: PDF of $\tau_\eta
    v^\parallel/r$ showing the $r$-scaling of the distribution bulk.
    (b) Same as (a) for $\St=0.3$ with $\beta$ = 2.2.  Inset: plots of
    $lnP_{-5}=\ln [\mathrm{Pr}\,(v^\parallel/u_{\eta}< 5 \,| r)]$ for
    $\St$ = 0.5 (blue), 0.3 (red), 0.2 (green) and 0.04 (gold
    crosses).  The values of $\beta$ were obtained as the linear slope
    of each of these curves, namely $\beta_{St=0.5}=2.1$,
    $\beta_{St=0.3}=2.2$ and $\beta_{St=0.2}=2.7$.  Unambiguous values
    of $\beta$ could not be obtained for $\St=0.04$.}
\end{figure}

The problem of droplet collision-coalescence in clouds involves
droplet relative velocities at contact, which is typically of the
order of 100 times smaller than $\eta$. \new{To that end, it is of
interest to understand how droplet relative velocities scale with
vanishing $r$ (granted that other inter-particle forces at small
scales will need to be accounted for a full description)}.
Figure~\ref{fig:jointpdfs}(a) presents the PDF of $v^\parallel$
conditioned on different values of $r$ for $\St=0.5$.  We find that
both the experimental and DNS data collapse at large negative values
of $v^\parallel$ when the PDF is rescaled by $r^\beta$ with $\beta
\approx 2.1$.  Similar analysis for the case of $St=0.3$ is shown in
Figure~\ref{fig:jointpdfs}(b).  Such collapse indicates that the
distribution of violent approaching velocities takes the form
$p(v^\parallel\,|\,r) \simeq r^{\beta(\St)}\,\phi(v^\parallel)$ at
sufficiently small separations and large
velocities\new{.\cite{Gustavsson2013}}
%This behavior is expected to
%extend down to separations of the order of the particle size and hence
%should describe the distribution of violent impact velocities between
%particles.  
It is straightforward to show analytically\cite{Celani2000} that the
exponent $\beta$ corresponds exactly to the saturated value
$\xi_{\infty}$ of the scaling exponents of the structure functions of
particle relative velocities in the limit of large
order\cite{bec2010intermittency} (i.e.,~$\langle
|v^\parallel|^p\,|\,r\rangle \propto r^{\xi_p}$ with $\xi_p =
\xi_{\infty}$ for all sufficiently large $p$).  The collapse to a
scale-independent form occurs for large velocity differences, namely
$|v^\parallel| \gg r/\tau_\mathrm{p}$.  This condition corresponds to
a traveling time over a distance $r$ that is much shorter than the
particle response time, so that damping is negligible.  Under these
conditions particle pairs move ballistically, which is related to the
sling
effect.\new{\cite{falkovich2002acceleration,Bewley2013,gustavsson2011distribution,Gustavsson2013}}
Gustavsson and Mehlig\cite{gustavsson2011distribution,Gustavsson2013}
predict that $\xi_{\infty}=3-D_2\equiv c_1$, where $D_2$ is the
fractal (correlation) dimension of inertial particle clusters and
$c_1$ the corresponding exponent of the radial distribution function.
In the case of $\St=0.5$, $\xi_{\infty} \simeq 0.7$ using value of
$c_1$ from e.g.\ Ref.~\onlinecite{Saw2012_p1}. \new{Our measured value
  $\xi_{\infty}(\St=0.5) \simeq 2.1$ differs from the prediction;
  this deviation could however disappear at much smaller $r$-scales.}

\begin{figure}
  %\subfigure[]{\includegraphics[width=0.3\textwidth]{bi-multifractal.pdf}}
  %\subfigure[]{\includegraphics[width=0.3\textwidth]{bi-multifractal_S_h.pdf}}
  %\subfigure[]{\includegraphics[width=0.3\textwidth]{Sh_vs_h_expr_dns_pub.pdf}}
  \vspace{-50pt}
  {\includegraphics[width=0.34\textwidth]{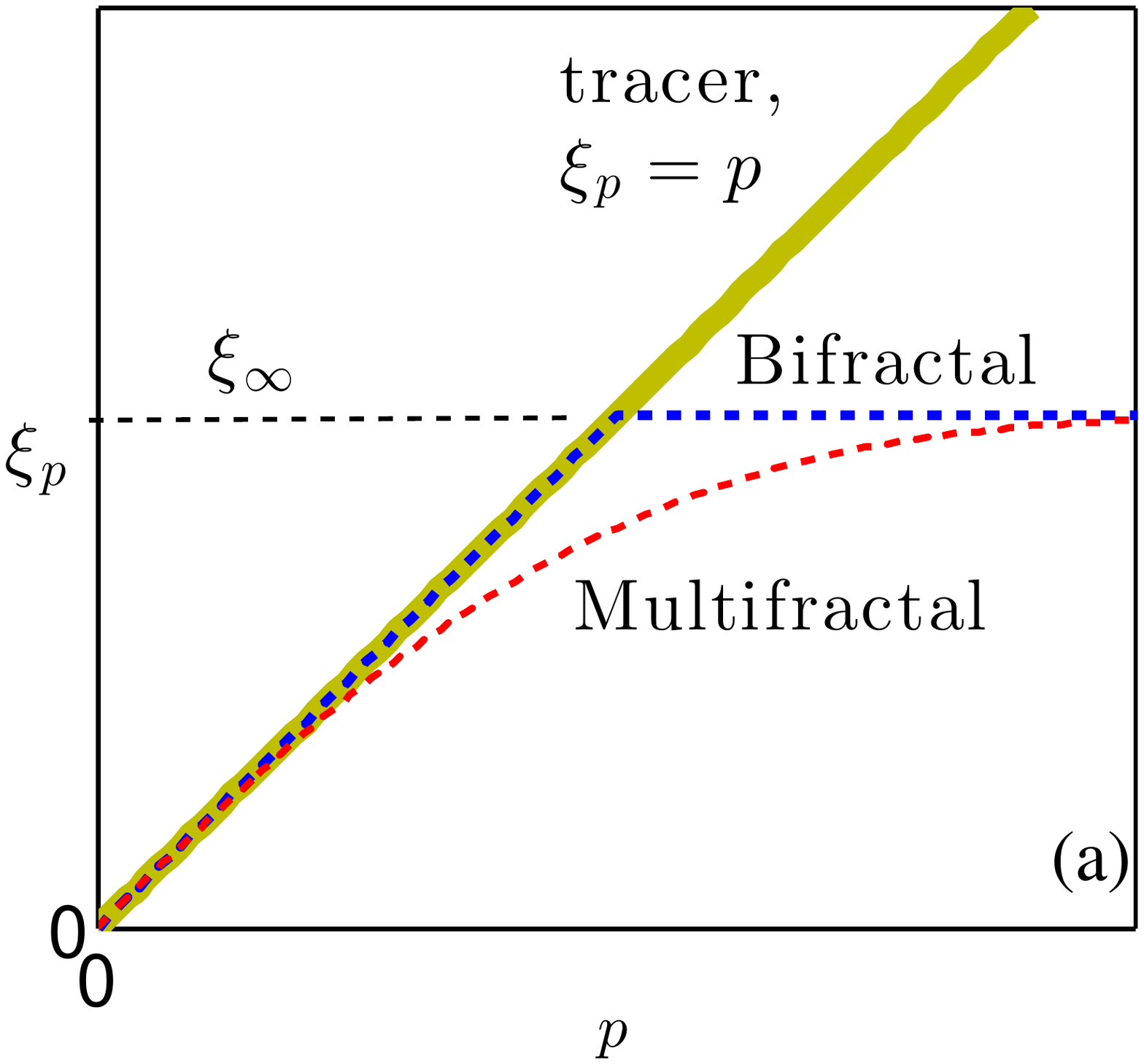}}
  \hspace{-15pt}
  {\includegraphics[width=0.34\textwidth]{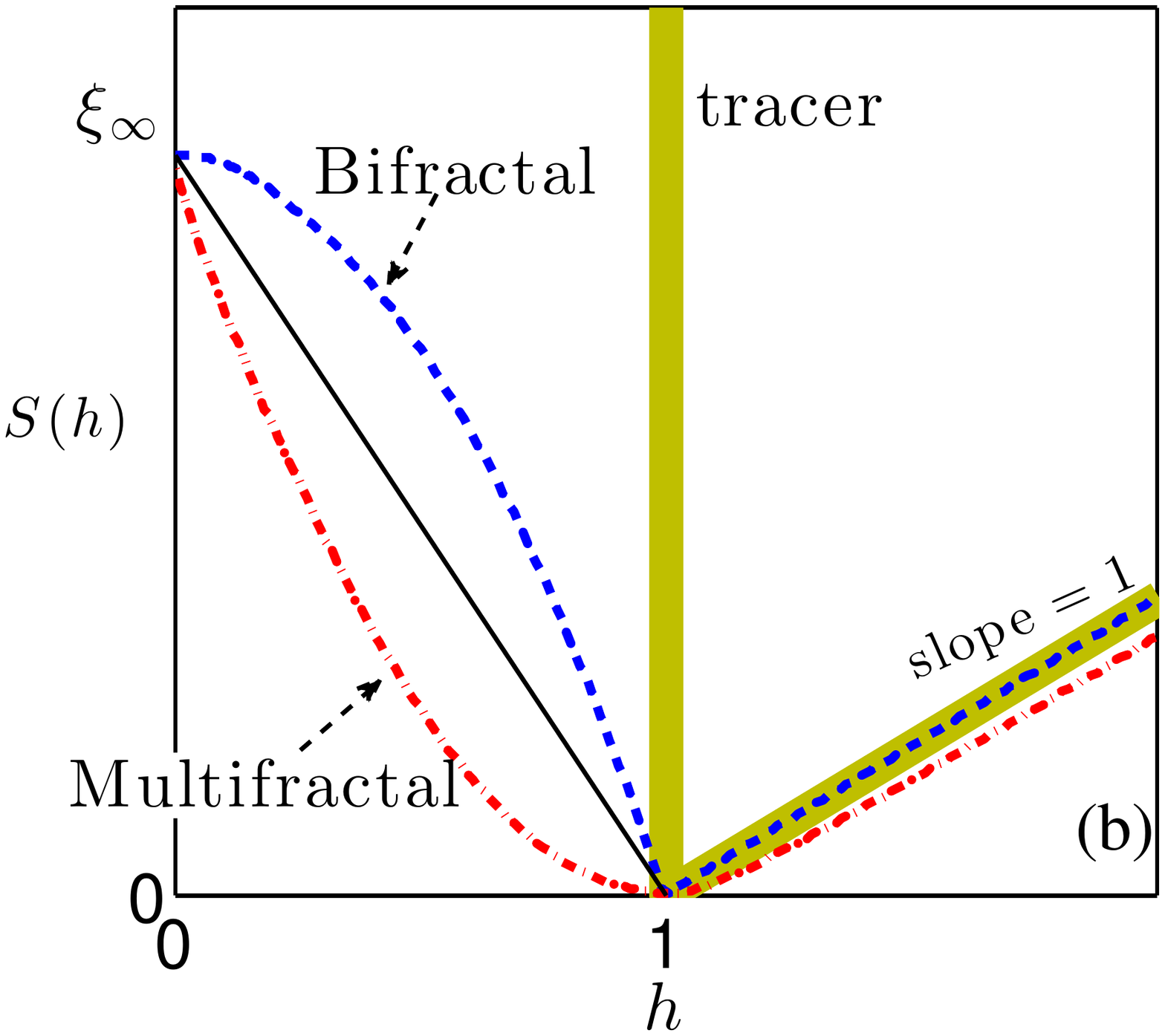}}
  \hspace{-15pt}
  {\includegraphics[width=0.34\textwidth]{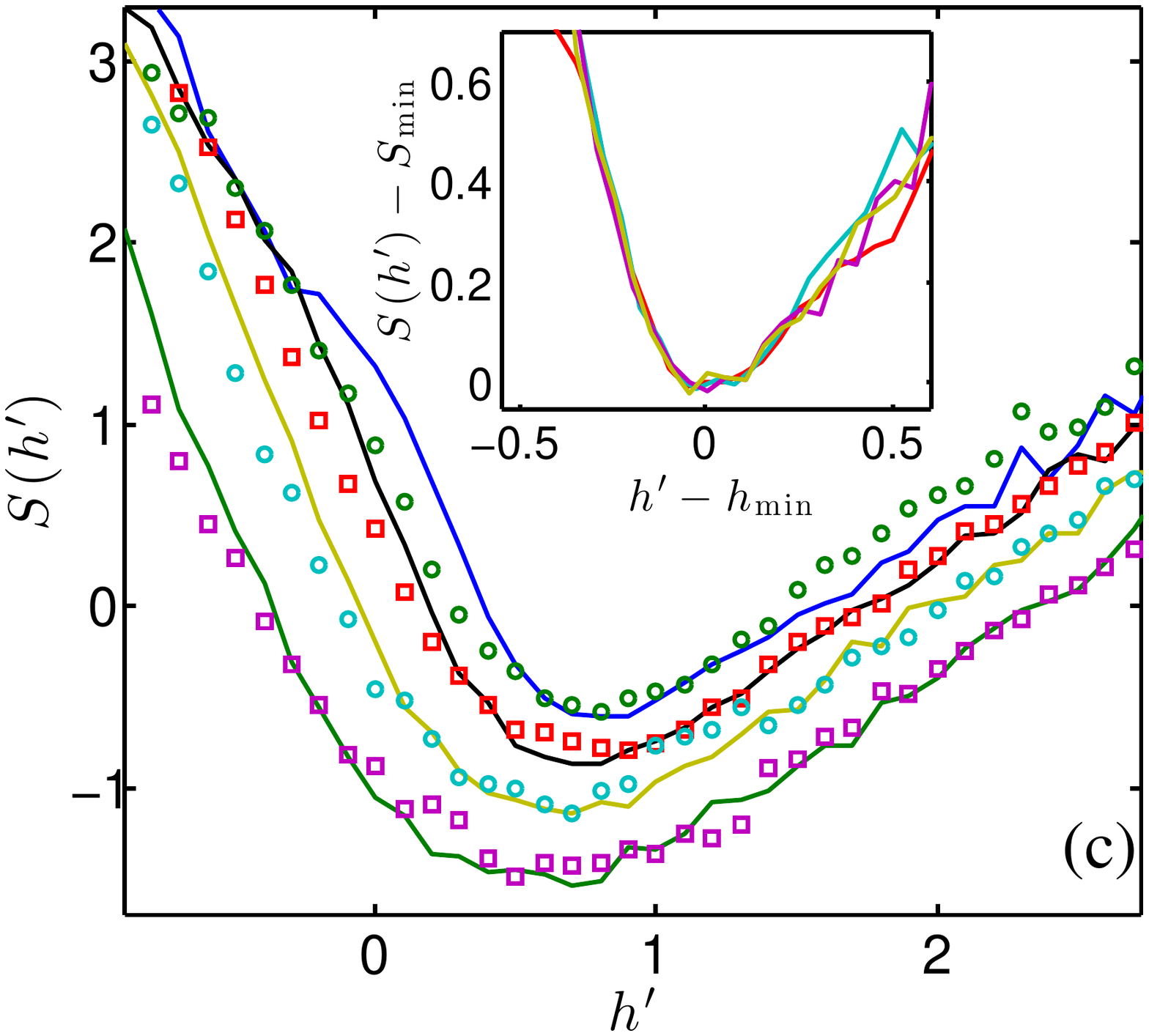}}
  \vspace{-45pt}
  \caption{\label{fig:bi-multi} (Color online) (a) and (b):
    \new{Schematic representation of} the main features of
    multi-fractal and bi-fractal statistics and the behaviors expected
    for fluid tracers in the dissipative scales of turbulent flows.
    (a) Scaling exponent $\xi_p$ (such that
    $\langle|v^\parallel|^p\,|\,r\rangle \propto r^{\xi_{p}}$).  (b)
    Corresponding rate functions $S(h)$ (see text for details).  (c)
    Rate function, $S(h)$, in the case of $\St=0.5$ for the experiment
    (markers) and the DNS (solid lines) and various
    separations. $S(h)$ was calculated using
    $S(h)=\ln[p(h\,|\,r_2)/p(h\,|\,r_1)]/\ln(r_2/r_1)$, where the
    $r_i$'s are the middle values of the $r$-bins with width $\Delta
    r=0.6\eta$.  From the bottom curve to the top, the values of
    $(r_1/\eta, r_2/\eta)$ were respectively (1.3, 2.3), (2.3, 3.3),
    (3.3, 4.3) and (4.3, 5.3).  Inset: Shifted $S(h)$ at smaller $r$'s
    using the DNS data.  All curves are linearly shifted such that
    their minimum is at $(0, 1)$.  Collapse implies that the limiting
    form is reached at these $r$ values which are, in the order
    red-cyan-purple-gold, respectively: $(r_1/\eta, r_2/\eta)=$(0.15,
    0.45), (0.45, 0.75), (0.75, 1.05), (1.05, 1.35), with bin widths
    of $\Delta r=0.3\eta$.  }
\end{figure}

% We turn our attention to the scaling exponents, $\xi_{p}$, of the
% relative velocity statistics, whose values for asymptotically large
% $p$ we reported above.
\new{By using matched asymptotics techniques, Gustavsson and
  Mehlig\cite{gustavsson2011distribution,Gustavsson2013} proposed to
  approximate the scaling exponents $\xi_{p}$ of relative velocities
  as}
%argued that the distribution of scaling exponents is bi-fractal, i.e.,
\begin{equation}
  % \xi_p= \left\{
  % \begin{array}{l l}
  %   p & \quad \text{$p\leq \xi_{\infty}$ }\\
  %   \xi_{\infty} & \quad \text{$p > \xi_{\infty}$}
  % \end{array} \right.
  \xi_p= p \quad\mbox{for}\ p\leq \xi_{\infty} \qquad\mbox{and} \qquad \xi_p=
  \xi_{\infty} \quad\mbox{for}\ p> \xi_{\infty}.
  \label{eq:bifractal}
\end{equation}
Thence in the limit of small $r/\eta$, the core of the PDF of
$v^\parallel$ inherits the scaling of the fluid tracers, namely
$p(v^\parallel | r) \sim r^{-1} \, \psi(v^\parallel / r) $, with a
transition at $| v^\parallel | \propto r/\tau_p$ to a scaling in the
tails of the form $p(v^\parallel | r)\sim
r^{\,\xi_{\infty}}\,\Phi(v^\parallel)$ which we discussed
above. \new{The behavior (\ref{eq:bifractal}) pertains to bi-fractal
  statistics.}  As illustrated in Figs.~\ref{fig:bi-multi}a and b,
this is a special case of multi-fractal statistics, which are
ubiquitous in turbulence.\cite{Frisch1996} For the problem of droplet
collisions in clouds, which depends on the first moments of the
relative particle velocity and concerns the moderate $\St$ studied
here, distinguishing between the two possibilities is of consequence,
since this is where the difference between the two is most
significant.
%Such difference is further substantiated by the fact 
%that they pertain to the exponents of velocity scaling.  
%Bi-fractality implies a specific $r$-scaling of the PDF of $v^\parallel$.  

% but there is a saturation exponent.  
% contrast this with the multi fractal nature of turbulence, where there is no saturation for large orders.  

Our data are consistent with the bifractal picture given above 
for both asymptotically large and small $v^\parallel$, 
as shown in the main plot of Fig.~\ref{fig:jointpdfs} for the scaling of the tail 
and in the inset for the scaling of the core.  
However, the behavior in the transition range ($|v^\parallel | \approx r/\tau_p$) 
differentiates a bifractal from a multifractal, 
and the sharpness of the transition is hard to judge from this figure.  
Hence we take a different approach as shown below.  

Multifractal analysis emerged in the context of strange
attractors\cite{paladin1987anomalous} and of the anomalous scaling
observed for inertial-range statistics in turbulence.\cite{Frisch1996}
Typical methods rely on box-counting, or on evaluating
moments and scaling exponents.\cite{meneveau1991multifractal} In the
specific case of inertial-particle velocity differences in the
dissipation range, measuring the scaling exponents $\xi_p$ as a
function of $p$ is particularly difficult as it relies on fitting data
to power-laws at scales where statistics deteriorate.  For that
reason, we use here the interpretation of multifractal statistics in
terms of the theory of large deviations.\cite{broniatowski2001self} We
assume a continuum of local scaling exponents $h = \ln |
v^\parallel/v_\ell | / \ln(r/\ell)$, where $\ell$ is a typical length
of convergence to the scaling regime and $v_\ell$ the associated
velocity.  In the asymptotics $r \ll \ell$, the probability density of
$h$ reads $p(h\,|\,r) \sim (r/\ell)^{S(h)}$, where $S(h)$ is the
\emph{rate function} (furthermore, $S(h) = 3-D(h)$, where $D(h)$ is
the \emph{multifractal spectrum}, that is the dimension of the set of
points where $v^\parallel \sim r^h$).  The scaling exponents trivially
relate to the rate function by a Legendre transform $\xi_p =
\inf_h[ph+S(h)]$.  Typical behaviors of $\xi_p$ and $S(h)$ are
sketched in Fig.~\ref{fig:bi-multi}a and b.  For tracers,
dissipation-range velocity differences are given by $v^\parallel =
r\,\partial_r u$, where $\partial_r u$ is the radial fluid
velocity gradient.  This leads to $\xi_p=p$ and $S(h)=h-1$ for $h \ge
1$, and $S(h) = \infty$ otherwise.  For bifractal
statistics, $\xi_p = \min(p,\xi_\infty)$ and $S(h) = h-1$ for $h \ge
1$, and $S(h)$ is a concave function for $h < 1$.  In the
multifractal case there must be no sharp transition at any $p$ or $h$ %at $p=1$
and $S(h)$ is a convex function around its minimum.  Distinguishing
between bifractal and multifractal statistics can thus be recast as an
investigation into whether $S(h)$ is convex or concave near its
minimum.

The measurement of $h$ and of $S(h)$ requires some attention 
because their definitions include the undetermined scales $\ell$ and $v_\ell$.  
Particular definitions of $\ell$ and $v_\ell$ 
do not alter the values of $h$ and $S(h)$ in the limit $r \to 0$, 
but we cannot reach this limit in practice. 
The explicit dependence of $S(h)$ on $\ell$ and $v_\ell$ can be eliminated 
by using the formula
$\ln[p(h\,|\,r_2)/p(h\,|\,r_1)]/\ln(r_2/r_1)$.  
%which converges to $S(h)$ when $r_1$ and $r_2 \to 0$.  
There is however no such stratagem to make $h$ independent of $\ell$ and $v_\ell$.  
A given choice, say $\ell'$ and $u_\ell'$, leads to a measurement of the scaling exponent 
$h' = \ln | v^\parallel/v_\ell' | /\ln(r/\ell')$ 
that for any finite $r$ differs from reference choices of $\ell$ and $v_\ell$ by 
$h'=h + [h\ln(\ell/\ell') +\ln(v_\ell/v_\ell')]/\ln(r/\ell')$.  
Thus, we must choose definitions for $\ell'$ and $u_\ell'$.  

The main panel of Fig.~\ref{fig:bi-multi}c shows $S(h')$ obtained from
our experiments and DNS for $\ell' = 10\eta$ and $u_\ell' = u_{\eta}$,
and with $r$ going from $\simeq 5\eta \text{ to } \eta$.  The $y$-axis
intercept for the case of $r \simeq \eta$, albeit noisy, gives roughly
the value deduced from Fig.~\ref{fig:jointpdfs}, namely
$\xi_\infty\approx 2.1$.  The location of the minimum shifts towards
$(1,0)$ as $r$ decreases.  The vertical displacement is partly due to
the normalization factor present in $p(h\,|\,r)$, which itself
involves some $r$ dependence,\cite{van1988generalized} and can be
compensated by subtracting from $S(h')$ the value
$S_{\min}=S(h'_{\min})$ of its minimum.  The origin of the horizontal
displacement can be twofold: it is either due to finite-$r$ deviations
from the limiting form of $S(h)$ or to a mismatch in the definition of
$h'$ due to our arbitrary choice of $\ell'$ and $v_\ell'$.  The DNS
data were consistent with $h'_{\min} \simeq 1 + C/\ln(r/\ell')$,
giving a strong support to the second scenario.

In order to probe the limiting form of $S$ close to its minimum at
vanishing $r$, we show in the inset of Fig.~\ref{fig:bi-multi}c the
rate functions $S(h)$ from the DNS with their minima translated to
$(0,0)$, for $r \simeq 0.15\eta$ to $1.05\eta$.  The excellent
collapse of these curves around their minima suggests that they have
reached their final limiting form at $r \lesssim \eta$.  The frozen
curvature around the minimum, for about a decade in $r$, indicates
that $S(h)$ is convex and thus supports the view that the statistics
are multi-fractal, and not bifractal. \new{However, we cannot rule out
  that this is an intermediate regime and that bi-fractality could be
  recovered at even smaller separations.}

To summarize, we evaluated the accuracy of the Stokes drag model for the advection of inertial particles 
in turbulent flow by comparing the results from DNS with experimental measurements.  
Focussing on large (longitudinal) relative velocities, 
we found that DNS reproduced all qualitative trends of the experiments.  
Furthermore, accurate quantitative agreements were found for inertia-dominated regimes 
($\St=0.5$, $v^{\parallel}\lesssim -r/\tau_\mathrm{p}$).  
Discrepancies up to a factor of 5 were found for regimes less influenced by particle inertia 
(that is, for separating particles or for small $\St$).  
Further analysis did not support trivial explanations for such discrepancies, 
which implies that the discrepancies could have been caused either 
by corrections to the Stokes drag model, 
such as the Basset history force or hydrodynamic interactions between particles, 
or by small-scale non-universality of the turbulence (DNS and experiment have different large scale energy injection schemes).  
Where the data agree, they consistently show that for inertial particles 
and at dissipative scales of turbulence, 
the tails of the probability density function of $v^{\parallel}$ 
scale as a power law of $r$.  
This is consistent with the saturation of the scaling exponents 
of the moments of velocities differences 
found in previous studies.  
Furthermore, the frozen convexity of the rate function, $S(h)$, 
at small $r$ is consistent with multi-fractal statistics of velocities differences.  

Several questions remain open.  \new{Foremost there is a clear need to resolve the velocity difference statistics at very small
  particle separations, in order to assess the recent theories (Ref.~\onlinecite{gustavsson2011distribution,Gustavsson2013}). Also, very
  little is known about the effect of turbulent intermittency on the
  statistics of caustics; this could lead to non-trivial Reynolds
  number dependencies of particle relative velocity and collision
  statistics. This would, for instance, make it possible to disentangle
  Reynolds number effects from Stokes number effects.}  These
questions will be addressed in future work.

We acknowledge Poh Yee Lim for help with the experiments; Holger
  Homann for help with the simulation; M.~Cencini, B.~Mehlig and
  S.~Musacchio for crucial discussions.  This work received funding
from the Max Planck Society (Germany) and the European Research
Council under the European Community's Seventh Framework Program
(FP7/2007-2013, Grant Agreement no. 240579).  SSR acknowledges the
support of the Indo-French Center for Applied Mathematics (IFCAM) and
from the AIRBUS Group Corporate Foundation Chair in Mathematics of Complex Systems established in ICTS.  Computations were performed on the IBM Blue Gene/P computer
JUGENE at the FZ J\"ulich was made available through the PRACE project
PRA031 and on the ``m\'esocentre de calcul SIGAMM''.

\bibliography{biblio_RVP}

\end{document}